\documentclass[prl,nofootinbib,floats,aps,twocolumn,tightenlines,superscriptaddress]{revtex4-1}
\usepackage[utf8]{inputenc}
\usepackage{amsmath}
\usepackage{amsfonts}
\usepackage{dsfont} 
\usepackage{graphicx}
\usepackage{todonotes}
\usepackage{gensymb}
\usepackage{bbold}
\usepackage{bm}
\usepackage{natbib}
\usepackage{amssymb}
\usepackage{latexsym}
\usepackage{fancyhdr}
\usepackage[bbgreekl]{mathbbol}
\usepackage{epsfig}
\usepackage{color}
\usepackage{textcomp}
\usepackage{amsthm}
\usepackage{comment}
\usepackage{ulem}
\usepackage{hyperref}

\usepackage{mathrsfs}

\newcommand{\Tr}[1]{\mathrm{Tr}\!\left(#1\right)}

\newcommand{\Id}{\mathds{1}}

\newcommand{\be}{\begin{equation}}
\newcommand{\ee}{\end{equation}}

\begin{document}

\title{Witnessing Opto-Mechanical Entanglement with Photon-Counting}
\date{\today}

\author{Melvyn Ho}
\affiliation{Quantum Optics Theory Group, University of Basel, Klingelbergstrasse 82, 4056 Basel, Switzerland}
\author{Enky Oudot}
\affiliation{Quantum Optics Theory Group, University of Basel, Klingelbergstrasse 82, 4056 Basel, Switzerland}
\author{Jean-Daniel Bancal}
\affiliation{Quantum Optics Theory Group, University of Basel, Klingelbergstrasse 82, 4056 Basel, Switzerland}
\author{Nicolas Sangouard}
\affiliation{Quantum Optics Theory Group, University of Basel, Klingelbergstrasse 82, 4056 Basel, Switzerland}

\begin{abstract}
The ability to coherently control mechanical systems with optical fields has made great strides over the past decade, and now includes the use of photon counting techniques to detect the non-classical nature of mechanical states. These techniques may soon be used to perform an opto-mechanical Bell test, hence highlighting the potential of cavity opto-mechanics for device-independent quantum information processing. Here, we propose a witness which reveals opto-mechanical entanglement without any constraint on the global detection efficiencies in a setup allowing one to test a Bell inequality. While our witness relies on a well-defined description and correct experimental calibration of the measurements, it does not need a detailed knowledge of the functioning of the opto-mechanical system. A feasibility study including dominant sources of noise and loss shows that it can readily be used to reveal opto-mechanical entanglement in present-day experiments with photonic crystal nanobeam resonators. 
\end{abstract}
\maketitle

\textit{Introduction.}--- Bell tests have initially been proposed to show that correlations between the results of measurements performed on two separated systems cannot be reproduced by classical strategies~\cite{Bell64}. They have been used to show the limit of classical physics as a complete description of small systems involving two atoms~\cite{Hensen15, Rosenfeld17} or two photons~\cite{Giustina15, Shalm15}. This naturally raises the question of a Bell inequality violation with larger systems. Concrete proposals have been made recently along this line to realise Bell tests with cavity opto- and electro-mechanical systems~\cite{Caprara16, Hofer16, Asjad18}. \\

Cavity opto-mechanics is at the core of intense research where the cavity field is used to control the motion of a mechanical system via radiation pressure. While initial efforts have focused on the cooling of mechanical oscillators down to the ground state~\cite{O'Connell10, Teufel11, Chan11}, impressive results including the detection of electro-~\cite{Palomaki13} and opto-mechanical~\cite{Riedinger16, Hong17} non-classical correlations and entanglement between two remote mechanical systems~\cite{Riedinger17} are now suggesting that cavity opto-mechanics could serve as a building block of future quantum networks~\cite{Kimble08} for the creation and storage of quantum information~\cite{Borkje11, Galland14}. If one is to show that cavity opto-mechanics can form the cornerstone of future quantum networks, it is crucial to prove that it is qualified for all possible uses of such networks. This means that the qualification must be device-independent~\cite{Scarani12}, that is, it cannot rely on a physical description of the actual implementation. A particular model using seemingly harmless assumptions, on the underlying Hilbert space dimension for instance, can completely corrupt the security guarantees that are offered by quantum networks for secure communications over long distances~\cite{Acin06,Lydersen10}. Device-independent schemes have been derived to certify all the building blocks of quantum networks that can be used to create, store or process quantum information~\cite{Sekatski18}. They could be directly implemented from the Bell tests proposed in Refs.~\cite{Caprara16, Hofer16}. Opto-mechanical Bell tests are thus not only of fundamental interest but are resources to certify the usefulness of opto-mechanical systems for long distance quantum communication with device-independent security guarantees. \\

\begin{figure}
\includegraphics[width=0.4\textwidth]{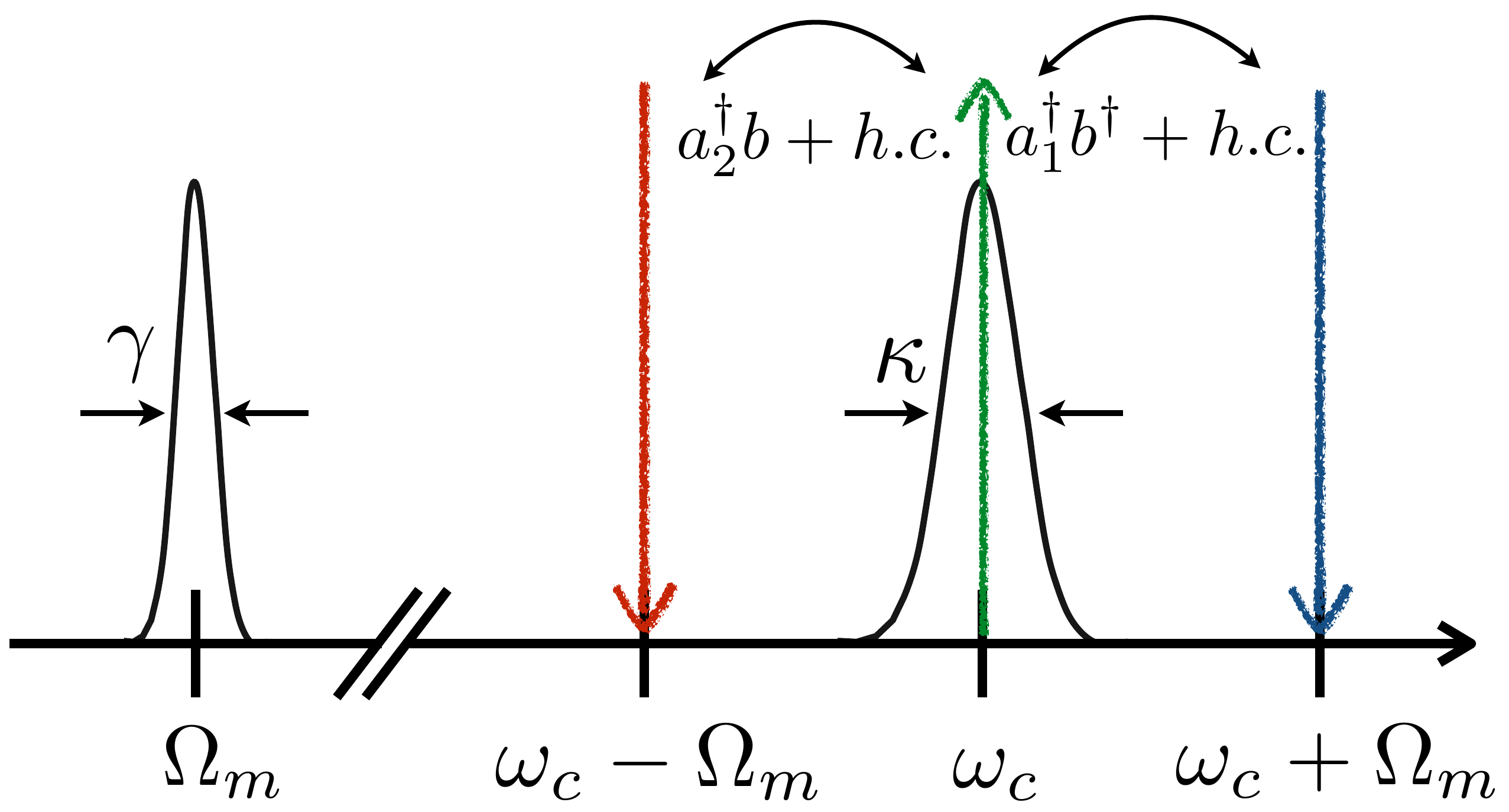}
\caption{A cavity optomechanical system is made with a cavity with frequency $\omega_c$ and a mechanical oscillator with frequency $\Omega_m.$ $\kappa$ and $\gamma$ are the cavity and mechanical decay rates respectively. We consider the resolved sideband regime where $\Omega_m \gg \kappa.$ Starting with a cooled mechanical system, the cavity opto-mechanical system is first driven by a laser resonant with the blue sideband. Photon-phonon pairs are created by means of an effective squeezing operation $a_1^\dag b^\dag + h.c.,$ the bosonic operators $a_1$ and $b$ corresponding to the cavity photons and mechanical phonons. The quantum nature of the correlations between the cavity photon number and the phonon number can be revealed by applying a laser resonant with the red sideband. This effectively maps the phononic state to a photonic state through a beamsplitter interaction $a_2^\dag b + h.c.$ The resulting photonic state involving two temporal modes $a_1$ and $a_2$ is detected with a photon detector supplemented with a displacement operation in phase space.}
\label{fig1}
\end{figure}

The violation of a Bell inequality as proposed in Refs.~\cite{Caprara16, Hofer16, Asjad18} is, however, not trivial. Ref.~\cite{Caprara16} uses a cavity opto-mechanical system in the resolved sideband regime where the mechanical frequency is larger than the cavity decay rate. Once cooled, the mechanical system is excited by laser light resonant with the blue sideband, see Fig. \ref{fig1}. Photons of the laser can decay into phonon - photon pairs, the photon being resonant with the cavity frequency and the phonon corresponding to a single excitation of the vibrational mode of the mechanical system. Energy conservation ensures that for each phononic excitation of the mechanical state, the cavity mode gets populated with a photonic excitation. These quantum correlations between phonon and photon numbers are strong enough to violate a Bell inequality~\cite{Caprara16, Hofer16}. The way to show this consists first in mapping the phononic excitations to cavity photons using laser light driving the red opto-mechanical sideband. This leads to a two-mode photonic state, where each mode can subsequently be detected with photon counting techniques preceded by displacement operations in phase space. By changing the amplitude and phase of the local displacements, the Bell-Clauser-Horne-Shimony-Holt (Bell-CHSH)~\cite{CHSH69} inequality can be violated as long as the global detection efficiency is higher than 67\%. While several experiments have been realized combining cavity opto-mechanics in the revolved-sideband regime and photon counting~\cite{Cohen15, Riedinger16, Hong17, Riedinger17}, the requirement on the efficiency remains very challenging to meet.\\

Here we propose the first step of an entire research program aiming to violate a Bell inequality with opto-mechanical systems, that is, we propose a witness for revealing opto-mechanical entanglement in the same scenario. In opposition to Bell tests, our witness assumes a detailed description and correct experimental calibration of measurements. This allows us to get rid of the requirement on the detection efficiency, even without any assumptions about the measured state. A feasibility study shows that our witness can readily be used to reveal opto-mechanical entanglement in present-day experiments with photonic crystal nanobeam resonators. \\

\textit{Temporal evolution of the cavity field and mechanical system}---  Let us recall the physics of optomechanical systems in the resolved sideband and weak coupling regime, which has been presented, at least partially, in various references~\cite{Aspelmeyer10, Hofer11, Vanner13, Galland14, Caprara16}. We consider the optical and mechanical modes of an opto-mechanical cavity with frequencies $\omega_c$ and $\Omega_m$ respectively. The bosonic operators associated to optical mode are called $a$ and $a^\dag$ while we use $b$ and $b^\dag$ for the mechanical mode. $g_0$ denotes the bare opto-mechanical coupling rate, $\kappa$ and $\gamma$ the cavity and mechanical decay rates. The cavity opto-mechanical system is laser driven on the lower or upper mechanical sideband with corresponding frequencies $\omega_\pm = \omega_c \pm \Omega_m.$ The laser powers are labelled $P_\pm$ respectively. The full Hamiltonian includes the uncoupled cavity and mechanical $\mathcal{H}_0 =  \hbar \omega_c a^\dag a + \hbar \Omega_m b^\dag b$ systems, the opto-mechanical coupling $-\hbar g_0 a^\dag a (b^\dag + b)$ and the coupling between the cavity mode and the driving laser $\hbar \left(s_{\pm}^* e^{i\omega_\pm t} a+ s_\pm e^{-i\omega_\pm t} a^\dag \right)$ with $|s_\pm| = \sqrt{\kappa P_\pm/\hbar \omega_{\pm}}.$ In the interaction picture with respect to $\mathcal{H}_0$ and focusing in the weak coupling $g_0 \ll \kappa$ and resolved-sideband $\kappa \ll \Omega_m$ regimes, the temporal evolution is given by a set of effective Langevin equations~\cite{Galland14}
\begin{equation}
\frac{da}{dt} = \frac{i}{\hbar} [\mathcal{H}_\pm,a] - \frac{\kappa}{2} a + \sqrt{\kappa} a_{\text{in}}, \quad \frac{db}{dt} = \frac{i}{\hbar} [\mathcal{H}_\pm,b] 
\end{equation}
with $\mathcal{H}_+ = -\hbar g_0 \sqrt{n_+} \left(a^\dag b^\dag + h.c.\right)$ and $\mathcal{H}_- = -\hbar g_0 \sqrt{n_-} \left(a^\dag b + h.c.\right)$ for a blue and red detuned driving laser respectively. $n_\pm = \frac{|s_\pm|^2}{\Omega_m^2 + \kappa^2/4}$ is the intra-cavity photon number. $a_{\text{in}}$ is the noise entering the cavity. The mechanical decay and corresponding thermal noise are neglected, that is, we focus on timescales smaller than the thermal decoherence time of the mechanical system $\frac{\hbar \omega_m}{ k_B T_{\text{bath}} \gamma}$ where $k_B T_{\text{bath}}$ is the Boltzmann energy.\\

\textit{Phonon-Photon correlations in the resolved sideband regime}--- Let us first focus on the initial step where a laser drives the upper sideband. We use the subscript $1$ for the cavity field operators corresponding to this initial step. We proceed with an adiabatic elimination of the cavity mode $\frac{da_1}{dt}=0$ that is, we consider a temporal evolution which is long compared to $\kappa^{-1}.$ Together with the input/ouput relation, that is, $a_{1,\text{out}} = -a_{1,\text{in}} + \sqrt{\kappa} a_1,$ we get 
\begin{equation}
a_{1,\text{out}} = a_{1,\text{in}} + i \sqrt{2 \tilde{g}_+} b^\dag, \quad \frac{db_1}{dt}= \tilde{g}_+ b + i \sqrt{2 \tilde{g}_+} a_{1,\text{in}}^\dag
\end{equation}
where $\tilde{g}_+ = \frac{2g_0^2 n_+}{\kappa}.$ Integrating the previous equations and introducing the temporal modes $A_{1,\text{in}/\text{out}} (t)=\sqrt{\frac{2 \tilde{g}_+}{\pm 1 \mp e^{\mp2\tilde{g}_+ t}}} \int_{0}^{t} dt'  e^{\mp\tilde{g}_+ t'} a_{1, \text{in/out}}(t')$ \cite{Hofer11} leads to $A_{1,\text{out}}(t) = e^{\tilde{g}_+ t} A_{1,\text{in}}(t) + i \sqrt{e^{2 \tilde{g}_+ t} -1} b^\dag(0),$ $b(t) = e^{\tilde{g}_+ t} b(0) + i \sqrt{e^{2\tilde{g}_+ t}-1} A^\dag_{1,\text{in}}(t).$ These two solutions can be written as $A_{1,\text{out}}(t) = U_1^\dag(t) A_{1,\text{in}} U_1(t)$ and $b(t) = U_1^\dag(t) b(0) U_1(t)$ where the propagator $U_1(t)$ is given by
\begin{align}
U_1(t) = & e^{i \sqrt{1-e^{-2 \tilde{g}_+ t}}  A^\dag_{1,\text{in}} b^\dag} e^{-\tilde{g}_+ t (A^\dag_{1,\text{in}}A^\dag_{1,\text{in}} + b^\dag b +1)}  \nonumber   \\
& \times  e^{i \sqrt{1-e^{-2 \tilde{g}_+ t}} A_{1,\text{in}} b}.
\end{align}
When $U_1(t)$ is applied on the vacuum, phonon-photon pairs are created where the phonon number equals the photon number, each of them following a thermal distribution with mean excitation number $e^{2 \tilde{g}_+ t}-1.$ These correlations between the phonon and photon numbers are strong enough to violate a Bell inequality, c.f. below. \\

\textit{Phonon-Photon correlations as the basis for a Bell inequality violation}--- Consider the case where a laser drives the lower sideband. We use the subscript $2$ for the cavity field operators corresponding to this second step. Following the line of thought developed in the previous paragraph while introducing $\tilde{g}_- = \frac{2 g_0^2 n_-}{\kappa}$, we can show that the cavity field and photon operators evolve according to the propagator~\cite{Galland14, Caprara16}
\begin{align}
U_2(t) =& e^{i\sqrt{e^{2\tilde{g}_- t}-1}A_{2,\text{in}}b^\dagger}e^{-\tilde{g}_-t (A_{2,\text{in}}^\dagger A_{2,\text{in}} - b^\dagger b)}  \nonumber \\
\label{U2propagator}
&\times e^{i\sqrt{e^{2\tilde{g}_- t}-1}A_{2,\text{in}}^\dagger b}.
\end{align}
This corresponds to a beamsplitter-type evolution, performing a conversion between the phononic and photonic modes with probability $1-e^{-2 \tilde{g}_- t}.$ In the limit $\tilde{g}_- t \rightarrow \infty,$ the phononic mode is perfectly mapped to the photonic mode $A_{2, \text{out}}$ and the phonon-photon correlations created in the first step are mapped to two temporal  photonic modes $A_{1,\text{out}}$ and $A_{2,\text{out}}.$ If both the cavity and mechanical system are in the vacuum, these two photonic temporal modes are 
described by a vacuum squeezed state $U_2(t)^{\tilde{g}_- t \rightarrow \infty} U_1(T_1) |0\rangle = e^{- \tilde{g}_+ T_1} e^{-\sqrt{1-e^{- 2\tilde{g}_+ T_1}}A^\dag_{1,\text{out}}A^\dag_{2,\text{out}}} |00\rangle.$ Refs.~\cite{Kuzmich00, Lee09, Brask12} have shown that such a state violates the Bell-CHSH inequality when it is measured with photon detection preceded by a displacement operation in phase space, the phase and amplitude being used to change the measurement setting. Ref.~\cite{Caprara16} showed that a minimum detection efficiency of $\sim$67\% is necessary to observe a violation of the Bell-CHSH inequality. This minimum detection efficiency even increases if the mechanical system is not in its ground state initially~\cite{Caprara16}. These efficiencies include all the loss from the cavity to the detector and are thus challenging to obtain in practice. We show in the following sections a way around this requirement which consists in replacing the Bell-CHSH inequality by a witness inequality, which assumes a physical description and correct experimental calibration of the measurement devices.\\

\textit{Photon counting preceded by a displacement operation}--- We focus on the setup described before, with which a Bell inequality is tested using photon detections preceded by a displacement operation $D(\alpha).$ Before presenting our entanglement witness, we first comment on such a measurement. We consider the realistic case where the photon detector does not resolve the photon number, that is, only two measurement results can be produced at each run. The first result corresponds to 
\textquotedblleft no-detection\textquotedblright and is modelled by a projection on the vacuum $|0\rangle\langle 0|$. The second possible result is a conclusive detection corresponding to the projection into the orthogonal subspace, that is, $\Id - |0\rangle\langle 0|.$ If we attribute the outcome $+1$ to a no-detection and $-1$ to a conclusive detection, the observable including the displacement operation is given by $\bbsigma_{\alpha} = D(\alpha)^\dag \left(2|0\rangle\langle 0| - \Id\right)D(\alpha).$ In the qubit subspace $\{|0\rangle, |1\rangle\},$ $\bbsigma_{0}$ corresponds exactly to the Pauli matrice $\sigma_z,$ that is, the outcome $+1$ $(-1)$ is associated to a projection into the state $|0\rangle$ $(|1\rangle).$ When $\alpha$ increases, the positive-operator valued measure (POVM) elements associated to outcomes $\pm 1$ get closer to projections in the $x-y$ plane of the Bloch sphere having $|0\rangle$ and $|1\rangle$ as north and south poles respectively~\cite{Caprara15}. For $\alpha=1,$ these POVM elements are projections along non-unit vectors pointing in the $x$ direction, while for $\alpha=i,$ they are noisy projections along the $y$ direction. This means that photon detection supplemented by a displacement operation performs noisy measurements in the qubit space $\{|0\rangle, |1\rangle\}$ whose direction in the Bloch sphere can be chosen by controlling the amplitude and phase of the displacement. \\

\textit{Witnessing phonon-photon correlations in a qubit subspace}--- In order to clarify on how to witness entanglement in two-mode squeezed vacuum using local observables $\bbsigma_{\alpha},$ we consider the state projection in the qubit subspace $1/\sqrt{1+|\epsilon|^2}(|00\rangle + \epsilon |11\rangle).$ The sum of relevant coherence terms $|00\rangle\langle 11| + |11\rangle \langle 00|$ can be measured using the ideal observable $M_{\text{ideal}} = \frac{1}{2\pi}\int (\cos \varphi \sigma_x + \sin \varphi \sigma_y) \otimes (\cos \varphi \sigma_x - \sin \varphi \sigma_y) d\varphi.$ Since separable states are i) non-negative states and ii) they stay non-negative under partial transposition \cite{Peres96, Horodecki96}, these coherence terms are bounded by $2 \min \{\sqrt{p(0,0) p(1,1)}, \sqrt{p(0,1) p(1,0)}\}$ for two-qubit separable states. $p(i,j)$ is the probability for having $i$ photons in mode $A_1$ and $j$ photons in $A_2.$ Any state $\rho$ such that $\Tr{M_{\text{ideal}}\rho} > 2 \min \{\sqrt{p(0,0) p(1,1)}, \sqrt{p(0,1) p(1,0)}\}$ is thus entangled. Since $p(0,1) = p(1,0) =0$ and $\Tr{M_{\text{ideal}}\rho} = 2 \text{Re}(\epsilon)/(1+|\epsilon|^2)$ for a state of the form $1/\sqrt{1+|\epsilon|^2}(|00\rangle + \epsilon |11\rangle),$ the witness observable $M_{\text{ideal}}$ has the potential to detect entanglement in two-mode squeezed vacuum, in the experimentally relevant regime where the squeezing is small $2\tilde{g}_+ T_1 \ll 1,$ that is, when the two-mode squeezed vacuum is well approximated by its projection in the qubit subspace. This suggests that a relevant witness observable for our purpose is 
\begin{equation}
\label{Witness_observable}
M(\alpha, \beta)= \int_0^{2\pi} \frac{d\phi}{2\pi}  \mathbf{U}_\phi^\dagger (\bbsigma_{\alpha} \otimes \bbsigma_{\beta}) \mathbf{U}_\phi
\end{equation}  
where the unitary $\mathbf{U}_\phi = e^{ i \phi A_1^\dagger A_1}\otimes e^{- i \phi A_2^\dagger A_2}$ is used to randomize the phase of displacements through the averaging over $\phi.$ Note that in Eq. \eqref{Witness_observable}, the amplitude of displacements is a free parameter. Further note that we are interested in revealing entanglement at the level of the detection. The non-unit efficiency of the detector can be seen as a loss operating on the state, i.e. the beamsplitter modelling the detector inefficiency acts before the displacement operation whose amplitude is changed accordingly, see Appendix A. This allows us to derive a  witness observable with unit efficiency detection and to include the detector efficiency at the end, see Appendix B. \\

\textit{Witnessing phonon-photon correlations without dimensionality restriction}--- Using the property of separable states which stay positive under partial transposition, we show in Appendix B that the maximum mean value $M(\alpha,\beta)$ can take if the measured state is separable is such that
\begin{equation}
\max_{\rho_{\text{sep}}}(M(\alpha,\beta) \rho_{\text{sep}}) \leq S^\star(\alpha, \beta)
\end{equation}
where $S^\star(\alpha, \beta)$ depends on some joint probabilities $p(i,j)$ for having $i$ photons in mode $A_1$ and $j$ photons in $A_2$ and the marginal probabilities $p(n_{A_1} \geq 2)$ and $p(n_{A_2} \geq 2)$ to have strictly more than one photon in mode $A_1$ and $A_2$ respectively. These probabilities are bounded in two steps in practice. In the first step, the probability $P(\pm1\pm1 |00)$ and $P(\mp 1 \mp 1 |00)$ of having $\pm 1$ for the outcomes of the detection of mode $A_1$ and $A_2$ without displacement ($\alpha=\beta=0$) are measured. They provide the following upper bounds $p(0,0) \leq P(+1+1|0,0),$ $p(0,1) \leq P(+1-1|0,0),$ $p(1,0) \leq P(-1+1|0,0)$ and $p(1,1) \leq P(-1-1|0,0).$  Second, two detectors after a 50/50 beamsplitter are used to measure the probability to get a twofold coincidence $P_{\text c} (A_{1/2})$ after the beamsplitter for both mode $A_1$ and $A_2.$ These coincidence probabilities provide the upper bounds on the missing elements, that is, $p(2,1) \leq p(n_{A_1} \geq 2) \leq 2 P_{\text c} (A_{1})$ and $p(1,2) \leq p(n_{A_2} \geq 2) \leq 2 P_{\text c} (A_{2}).$ This results in a bound $S^\star(\alpha, \beta)$ whose value depends on the local displacement amplitudes $\alpha$ and $\beta.$ Finally, the  
mean value $Q(\alpha,\beta)$ of $M(\alpha, \beta)$ is measured by evaluating $P(+1+1|\alpha,\beta),$ $P(+1|\alpha)$ and $P(+1|\beta),$ that is 
\begin{equation}
\nonumber
Q(\alpha, \beta) = 1 - 2P(+1|\alpha) - 2P(+1|\beta)+ 4 P(+1+1|\alpha, \beta).
\end{equation}
If there is a value for the couple $\alpha, \beta$ such that $Q(\alpha,\beta) - S^\star(\alpha, \beta) >0,$ we deduce that the photonic modes $A_1$ and $A_2$ are entangled. Since the state describing $A_2$ is obtained from a local operation on the phononic state, $Q(\alpha,\beta) - S^\star(\alpha, \beta) >0$ also certifies photon-phonon entanglement.\\ 

\begin{figure}
\includegraphics[width=8 cm]{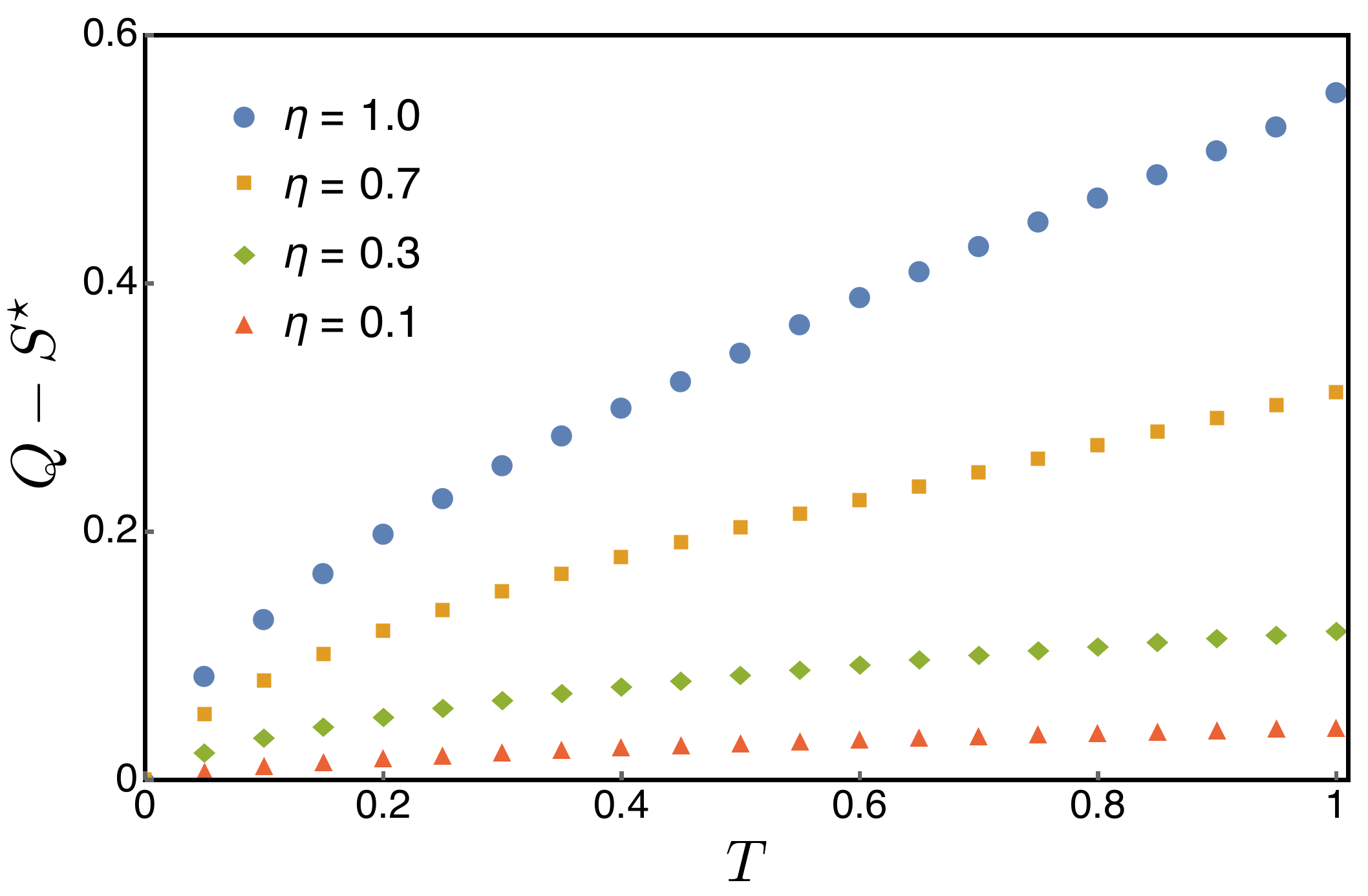}
\caption{Difference $Q - S^\star$ between the mean value of our witness observable $M(\alpha, \beta)$ that would be observed between the optical modes $A_1$ and $A_2$ and the maximum value that would be obtained with a separable state as a function of the phonon-photon conversion efficiency $T = 1 - e^{-2\bar{g}_- T_2}$ for various overall detection efficiencies $\eta$, optimised over displacement choices $\alpha, \beta$ and the amount of initial squeezing $\bar{g}_+ T_1$ which is kept small. $Q - S^\star>0$ witnesses entanglement.}
\label{Fig2}
\end{figure}

\textit{Results}--- We focus on the statistics that would be collected in modes $A_1$ and $A_2$ if the upper sideband is laser driven during the time interval $T_1$ and the lower sideband is subsequently driven for a duration $T_2.$ The value $Q - S^\star$ that would be obtained in this case when optimizing the arguments of local displacements $\alpha, \beta$ and the amount of initial squeezing $\bar{g}_+ T_1$ is shown in Fig. \eqref{Fig2} as a function of the phonon-photon conversion efficiency $T = 1 - e^{-2\bar{g}_- T_2}$ for various overall detection efficiency $\eta,$ see Appendix C for more details. Fig. \eqref{Fig2} shows a very favorable robustness of our witness to inefficiencies. We stress that the efficiency $\eta$ represents the global detection efficiency, including all the loss from the cavity optomechanical system to the detector. We here assumed that the mechanical system is prepared in its ground state. In the more realistic case where the initial mechanical cooling leads to a mechanical thermal state with non-zero mean occupation number $n_0,$ the results presented in Fig. \eqref{Fig2} for $\eta=0.3$ for example are essentially unchanged as long as $n_0 \leq 0.1$ and substantial differences between $Q$ and $S^\star$ can still be observed for $n_0 \sim1,$ see Appendix C.\\

\textit{Feasibility Study}--- To illustrate the feasibility, we focus on a photonic crystal nanobeam resonator~\cite{Chan11, Chan12, Kuramochi10} which distinguishes itself by a high mechanical frequency $\Omega_m/2\pi = 5.25$ GHz~\cite{Hong17}. Together with the cavity decay rate $\kappa/2\pi = 846$ MHz~\cite{Hong17} and the optomechanical coupling rate $g_0/2 \pi =  869$ kHz~\cite{Hong17}, this resonator is placed in the deep resolved sideband and weak coupling regimes. To control the initial number of excitations, we consider the use of a dilution refrigerator, which can bring the mean phonon number $n_0 \sim 0.2.$ Furthermore, to prevent decoherence of the phonon state we also consider pulse durations much smaller than the typical decoherence time of the oscillator, which is of the order of $10 \mu$s~\cite{Chan12, Sun13}. Considering a global detection efficiency $\eta = 10\%,$ an initial mean phonon number of $n_0 =0.2$ and state-swap efficiency of $T = 30\%$ which can be realised using a pulse laser resonant with the red sideband with a duration of $T_2 = 50$ns and intra-cavity photon number $n_-  \approx 318,$ we expect to conclude about the presence of entanglement (violation of the inequality $Q-S^\star <0$ by 3 standard deviations) within $750 000$ experimental runs, see Appendix D. This involves the creation of a phonon-photon state using a blue-detuned pulse of duration $T_1 = 50$ns and $n_+ \approx 298,$ and the choice of displacement amplitudes $\alpha= -\beta=2.63$. Given the experiments reported in Refs.~\cite{Hong17, Riedinger17}, we conclude that our scheme appears feasible with currently available technologies. \\

\textit{Conclusion}--- We have presented a witness tailored for the detection of opto-mechanical entanglement using photon countings. 
Our proposal is based on the measurement of single and twofold coincidence counts. It requires basic phase stabilizations and is robust to loss, see Appendix E. This makes us confident that it can be used in present day experiments with photonic crystal nanobeam resonators to show opto-mechanical entanglement. Following the proposal of Ref. \cite{Hofer16}, it also applies straightforwardly to electro-mechanical systems where it could be used to demonstrate electro-mechanical entanglement with non-gaussian resources. \\


\begin{acknowledgments}
We thank C. Galland and B. Gouraud for enlightening discussions. This work was supported by the Swiss National Science Foundation (SNSF), through the Grants PP00P2-150579 and PP00P2-179109. We also acknowledge the Army Research Laboratory Center for Distributed Quantum Information via the project SciNet.\\
\end{acknowledgments}

\bibliographystyle{unsrt}

\begin{appendix}

\section{Appendix A - Modelling detectors with non-unit efficiencies}
The proposed entanglement witness relies on measurements that are realized with non-photon resolving detectors preceded by displacement operations in phase space. As explained in the main text, we assign the outcome $+1$ to a no-detection and $-1$ to a conclusive detection. Given a state $\rho$ in the mode corresponding to the bosonic operators $A_1$ and $A_1^\dag,$ the probability to get the outcome $+1$ using a displacement with argument $\alpha$ is given by 
\begin{equation}
P(+1| \alpha) = \Tr{D(\alpha)^\dag |0\rangle\langle 0| D(\alpha) \rho}.
\end{equation}
So far, we assumed that the detector has unit efficiency. To model the detector inefficiency, a beamsplitter with transmission $\eta  = \cos^2 \theta$ can be introduced, that is
$$
P(+1| \alpha) = \Tr{D(\alpha)^\dag U_{\theta}^\dag |\bar 0\rangle\langle \bar 0| U_{\theta} D(\alpha) \rho}.
$$
with $U_{A_1c}=e^{\theta(A_1^\dag c - A_1 c^\dag)},$ the auxiliary mode described by $c$ and $c^\dag$ being initially empty. The state $|\overline{0} \rangle$ corresponds to the projection onto the vacuum for both $A_1$ and $c.$ Commuting the beamsplitter and displacement operation leads to   
$$
P(+1| \alpha) = \Tr{D(\sqrt{\eta} \alpha)^\dag |0\rangle\langle 0| D(\sqrt{\eta} \alpha) U_{\theta} \rho \otimes |0\rangle\langle 0| U_{\theta}^\dag}.
$$
This means that we can model the detection inefficiency as loss operating on the state that is measured if the amplitude of the displacement operation is changed accordingly. Hence, we consider detectors with unit efficiencies to derive our entanglement witness, and only replace the displacement amplitudes $\alpha \rightarrow \sqrt{\eta}\alpha,$ $\beta \rightarrow \sqrt{\eta}\beta$ at the end to account for the non-unit detection efficiency.

\section{Appendix B - Maximum value of the witness observable for separable states}
Our aim is to bound the value of the witness observable 
\begin{equation}
M(\alpha, \beta) = \int_0^{2\pi} \frac{d\phi}{2\pi}  \mathbf{U}_\phi^\dagger (\bbsigma_{\alpha} \otimes \bbsigma_{\beta}) \mathbf{U}_\phi
\end{equation}
with 
$$
\mathbf{U}_\phi = U^{A_1}_{\phi} \otimes U^{A_2}_{-\phi} = e^{i\phi A_1^\dag A_1} \otimes e^{-i\phi A_2^\dag A_2}
$$ 
when applied on separable states. We first use the fact that the trace is cyclic. Hence, the phase averaging can be applied on the state, that is
\begin{eqnarray}
\nonumber
&\Tr{M(\alpha, \beta) \rho} & = \Tr{\int_0^{2\pi} \frac{d\phi}{2\pi} (\bbsigma_{\alpha} \otimes \bbsigma_{\beta}) \mathbf{U}_\phi  \rho \mathbf{U}_\phi^\dagger } \\
\nonumber
&&= \int_0^{2\pi} \frac{d\phi}{2\pi}  \Tr{(\bbsigma_{\alpha} \otimes \bbsigma_{\beta})\mathbf{U}_\phi  \rho \mathbf{U}_\phi^\dagger },
\end{eqnarray}
where the last equality holds by linearity of the trace. Next, we recognise that the expectation value can be obtained from the partial transposed quantities if $\rho$ is separable, that is, for $\rho^{\text{sep}} = p_i \sum_i \rho^i_{A_1} \otimes \rho^i_{A_2},$ we have
\begin{eqnarray}
\nonumber
&\Tr{M(\alpha, \beta) \rho^{\text{sep}}} = \int_0^{2\pi} \frac{d\phi}{2\pi} &  \sum_i p_i \Tr{\bbsigma_\alpha^{T} (U^{A_1}_{\phi}  \rho^i_{A_1} U^{A_1\dag}_\phi)^T} \\
\label{witness_partial_transpose}
&&\times \Tr{\bbsigma_\beta U^{A_2}_{-\phi}  \rho^i_{A_2} U^{A_2\dag}_{-\phi}} 
\end{eqnarray}
where $T$ indicates the transpose in the photon number basis. This can be shown in the following way  
\begin{eqnarray}
\nonumber
&& \Tr{(\bbsigma_\alpha\otimes\bbsigma_\beta) \mathbf{U}_\phi  \rho^{\text{sep}} \mathbf{U}_\phi^\dagger}  \\
\nonumber
&& =  \Tr{(\bbsigma_\alpha\otimes\bbsigma_\beta) \mathbf{U}_\phi  \sum_i p_i \rho^i_{A_2} \otimes \rho^i_{A_1} \mathbf{U}_\phi^\dagger}\\
\nonumber
&& = \sum_i p_i \Tr{\bbsigma_\alpha U^{A_1}_{\phi}  \rho^i_{A_1} U^{A_1\dag}_\phi} \Tr{\bbsigma_\beta U^{A_2}_{-\phi}  \rho^i_{A_2} U^{A_2\dag}_{-\phi}} \\
\nonumber
&& = \sum_i p_i \Tr{\bbsigma_\alpha^{T} (U^{A_1}_{\phi}  \rho^i_{A_1} U^{A_1\dag}_\phi)^T} \Tr{\bbsigma_\beta U^{A_2}_{-\phi}  \rho^i_{A_2} U^{A_2\dag}_{-\phi}}. 
\end{eqnarray}
The previous expression can be further simplified using the properties of the transpose, that is
\begin{eqnarray}
\nonumber
&& \Tr{(\bbsigma_\alpha\otimes\bbsigma_\beta) \mathbf{U}_\phi  \rho^{\text{sep}} \mathbf{U}_\phi^\dagger}  \\
\nonumber
&& = \sum_i p_i \Tr{\bbsigma_\alpha^{T} (U^{A_1 \dag}_{\phi})^T  \rho^{i,T}_{A_1} (U^{A_1}_\phi)^T} \Tr{\bbsigma_\beta U^{A_2}_{-\phi}  \rho^i_{A_2} U^{A_2\dag}_{-\phi}} \\
\nonumber
&& = \sum_i p_i \Tr{\bbsigma_\alpha^{T} U^{A_1}_{-\phi}  \rho^{i,T}_{A_1} U^{A_1 \dag}_{-\phi}} \Tr{\bbsigma_\beta U^{A_2}_{-\phi}  \rho^i_{A_2} U^{A_2\dag}_{-\phi}}.
\end{eqnarray}
Further note that $\bbsigma_\alpha^{T}= D(\alpha^*)^\dag \left(2|0\rangle\langle 0| - \Id\right)D(\alpha^*)=\bbsigma_{\alpha^*}.$ Hence, 
\begin{eqnarray}
\nonumber
&& \Tr{(\bbsigma_\alpha\otimes\bbsigma_\beta) \mathbf{U}_\phi  \rho^{\text{sep}} \mathbf{U}_\phi^\dagger}  \\
\nonumber
&& = \sum_i p_i \Tr{\bbsigma_{\alpha^*} U^{A_1}_{-\phi}  \rho^{i,T}_{A_1} U^{A_1 \dag}_{-\phi}} \Tr{\bbsigma_\beta U^{A_2}_{-\phi}  \rho^i_{A_2} U^{A_2\dag}_{-\phi}} \\
\nonumber
&& = \Tr{(\bbsigma_{\alpha^*} \otimes\bbsigma_\beta) (U^{A_1}_{-\phi}\otimes U^{A_2}_{-\phi})  \rho^{\text{sep}, T_{A_1}}  (U^{A_1 \dag}_{-\phi}\otimes U^{A_2 \dag}_{-\phi})}. 
\end{eqnarray}
Therefore 
\begin{equation}
\Tr{M(\alpha, \beta) \rho^{\text{sep}}} = \Tr{(\bbsigma_{\alpha^*} \otimes\bbsigma_\beta) \rho^{\text{sep},T _{A_1}}_{\text{rand}}}
\end{equation}
where 
\begin{equation}
\nonumber
\rho^{\text{sep},T _{A_1}}_{\text{rand}} = \int_0^{2\pi} \frac{d\phi}{2\pi} (U^{A_1}_{-\phi}\otimes U^{A_2}_{-\phi})  \rho^{\text{sep}, T_{A_1}}  (U^{A_1 \dag}_{-\phi}\otimes U^{A_2 \dag}_{-\phi}).
\end{equation}
Interestingly, $\rho^{\text{sep},T _{A_1}}_{\text{rand}}$ has a simple structure due to the phase randomization. It can be written as   
\begin{eqnarray}
\rho^{\text{sep},T _{A_1}}_{\text{rand}}= 
\left[
\begin{array}{c c c c| c c c c c} 
    \checkmark & & & & & & & & \\ 
    &\checkmark&\checkmark& & & & & & \\
    &\checkmark&\checkmark& & & & & & \\
    & & &\checkmark&\checkmark&\checkmark& & & \\ \hline
    & & &\checkmark&.&.&.&.&  \\ 
    & & &\checkmark&.&.&.&.& \\ 
    & & &&.&.&.&.& 
\end{array}
\right]
\end{eqnarray}
in the basis $\{|00\rangle, |01\rangle, |10\rangle, |11\rangle, |02\rangle, |20\rangle...\}.$ The checkmarks indicate non-zero terms. The upper block on the left corresponds to the projection in the qubit subspace where modes $A_1$ and $A_2$ are filled with at most one photon each. Similarly, the lower block on the right corresponds to the projection on a subspace where at least one mode is filled with at least two photons. The anti-diagonal blocks correspond to coherences between these two subspaces. Considering the contributions from each of these blocks separately, we obtain 
\begin{align}
\nonumber
&\Tr{M(\alpha, \beta) \rho^{\text{sep}}} =  \Tr{(\bbsigma_{\alpha^*} \otimes \bbsigma_{\beta}) \rho_{\text{rand}}^{\text{sep},T_{A_1}, n_{A_1}\leq 1 \cap n_{A_2}\leq 1}} \\
\nonumber
&+ 2  \times \text{Re} [\langle 11|(\bbsigma_{\alpha^*} \otimes \bbsigma_{\beta}) |02 \rangle \langle 02|\rho_{\text{rand}}^{\text{sep}, T_{A_1}} | 11\rangle]\\
\nonumber
&+ 2\times  \text{Re} [\langle 11|(\bbsigma_{\alpha^*} \otimes \bbsigma_{\beta}) | 20 \rangle \langle 20|\rho_{\text{rand}}^{\text{sep},T_{A_1}}| 11\rangle]\\
\label{inter_sep}
&+\Tr{(\bbsigma_{\alpha^*} \otimes \bbsigma_{\beta}) \rho_{\text{rand}}^{\text{sep},T_{A_1}, n_{A_1}\geq 2 \cup n_{A_2}\geq 2}}
\end{align}
where $\rho_{\text{rand}}^{\text{sep},T_{A_1}, n_{A_1}\leq 1 \cap n_{A_2}\leq 1}$ corresponds to the projection in the qubit subspace and $\rho_{\text{rand}}^{\text{sep},T_{A_1}, n_{A_1}\geq 2 \cup n_{A_2}\geq 2}$ is the state projection in the subspace with two photons or more in at least one of the modes.\\

Eq. \eqref{inter_sep} allows us to bound the value of the witness observable in the case where the measured state is separable. The reasoning is the following: Let $p(i,j)$ be the probability of having $i$ photons in $A_1$ and $j$ photons in $A_2$ which is a diagonal element of the measured state in the Fock basis. 
If this state is separable, we have $p(i,j) = \langle i,j|\rho^{\text{sep}} |i,j \rangle = \langle i,j | \rho_{\text{rand}}^{\text{sep},T_{A_1}} | i, j\rangle.$ Furthermore $|\langle i ,j+1 | \rho_{\text{rand}}^{\text{sep},T_{A_1}} | i+1 ,j \rangle| = |\langle i+1, j+1 | \rho^{\text{sep}} | i,j \rangle| \leq \min\{\sqrt{p(i,j)p(i+1,j+1)},\sqrt{p(i,j+1)p(i+1,j)}\}$ since both $\rho^{\text{sep}}$ and its partial transpose are positive. Similarly, we have $|\langle i+1 ,j | \rho_{\text{rand}}^{\text{sep},T_{A_1}} | i, j+1 \rangle| = |\langle i, j | \rho^{\text{sep}} | i+1,j+1\rangle| \leq \min\{\sqrt{p(i,j)p(i+1,j+1)},\sqrt{p(i+1,j)p(i,j+1)}\}.$ 
This means that the value of the first term in Eq. \eqref{inter_sep} is bounded by   
\begin{eqnarray}
\nonumber
&& \Tr{(\bbsigma_{\alpha^*} \otimes \bbsigma_{\beta}) \rho_{\text{rand}}^{\text{sep},T_{A_1}, n_{A_1}\leq 1 \cap n_{A_2}\leq 1}} \\
\nonumber
&& =  \sum_{i,j=0}^1  \langle ij|\bbsigma_{\alpha^*} \otimes \bbsigma_\beta |ij \rangle p(i,j) \\
\nonumber
&&  \quad  \quad + 2 \times \text{Re}[\langle 01|\bbsigma_{\alpha^*} \otimes \bbsigma_\beta|10 \rangle  \langle 00|  \rho^{\text{sep}} |11 \rangle ] \\
\nonumber
&&  
\leq   \sum_{i,j=0}^1  \langle ij|\bbsigma_{\alpha^*} \otimes \bbsigma_\beta |ij \rangle p(i,j) \\
\nonumber
&&  \quad  \quad + 2 |\langle 01|\bbsigma_{\alpha^*} \otimes \bbsigma_\beta|10 \rangle  \langle 00|  \rho^{\text{sep}} |11 \rangle| \\
\nonumber
&&  
=  \sum_{i,j=0}^1  \langle ij|\bbsigma_{\alpha^*} \otimes \bbsigma_\beta |ij \rangle p(i,j) \\
\nonumber
&&  \quad  \quad + 2 |\langle 01|\bbsigma_{\alpha^*} \otimes \bbsigma_\beta|10 \rangle | |\langle 00|  \rho^{\text{sep}} |11 \rangle| \\
\nonumber
&& \leq  \sum_{i,j=0}^1  \langle ij|\bbsigma_{\alpha} \otimes \bbsigma_\beta |ij \rangle p(i,j) \\
\nonumber
&&  \quad  \quad + 2 |\langle 01|\bbsigma_{\alpha} \otimes \bbsigma_\beta|10 \rangle| \\
\nonumber
&& \quad  \quad \quad \quad \times
\min\{\sqrt{p(0,0)p(1,1)},\sqrt{p(0,1)p(1,0)}\}
\end{eqnarray}
where in the last 2 lines, $\alpha$ and $\beta$ can be considered as real numbers without loss of generality.
Similarly, the coherences in the second and third terms are bounded by 
\begin{eqnarray}
\nonumber
&& 2 \times \text{Re}[\langle 11|\bbsigma_{\alpha^*} \otimes \bbsigma_{\beta} |02 \rangle \langle 02| 
\rho_{\text{rand}}^{\text{sep}, T_{A_1}} | 11\rangle] \\
\nonumber
&& \quad \leq 2 |\langle 11|\bbsigma_{\alpha} \otimes \bbsigma_{\beta} |02 \rangle| \\
\nonumber
&& \quad \quad \times \min\{\sqrt{p(1,2)p(0,1)}, \sqrt{p(0,2)p(1,1)}\}
\end{eqnarray}
and
\begin{eqnarray}
\nonumber
&&2 \times \text{Re}[\langle 11|\bbsigma_{\alpha^*} \otimes \bbsigma_{\beta} | 20 \rangle \langle 20|\rho_{\text{rand}}^{\text{sep},T_{A_1}}| 11\rangle]\\
\nonumber
&& \quad \leq 2 |\langle 11|\bbsigma_{\alpha} \otimes \bbsigma_{\beta} |20 \rangle|\\
\nonumber
&& \quad \quad \times \min \{\sqrt{p(1,0)p(2,1)}, \sqrt{p(2,0)p(1,1)} \}.
\end{eqnarray}
As for the last term, we use the fact $\rho^{\text{sep},T _{A_1}}_{\text{rand}}$ is a physical state, so that its projection into the subspace where there is at least two photons in at least one mode is also a physical state with a norm given by $\Tr{\rho_{\text{rand}}^{\text{sep},T_{A_1}, n_{A_1}\geq 2 \cup n_{A_2}\geq 2}} = p(n_{A_1}\geq 2 \cup n_{A_2}\geq 2).$ The maximum eigenvalue of the observable $(\bbsigma_{\alpha^*} \otimes \bbsigma_{\beta})$ being one, we conclude that 
\begin{eqnarray}
\nonumber
&&\Tr{(\bbsigma_{\alpha^*} \otimes \bbsigma_{\beta}) \rho_{\text{rand}}^{\text{sep},T_{A_1}, n_{A_1}\geq 2 \cup n_{A_2}\geq 2}}\\
&& \quad \leq p(n_{A_1} \geq 2) + p(n_{A_2} \geq 2).
\end{eqnarray}
Hence, a bound on the maximum mean value $S^\star(\alpha, \beta)$ that $M(\alpha,\beta)$ can take if the measured state is separable 
\begin{equation}
\max_{\rho_{\text{sep}}}(\Tr{M(\alpha,\beta) \rho_{\text{sep}}}) \leq S^\star(\alpha, \beta)
\end{equation}
can be obtained by upper bounding some joint probabilities $p(i,j)$ for having $i$ photons in mode $A_1$ and $j$ photons in $A_2$ and the marginal probabilities $p(n_{A_1} \geq 2)$ and $p(n_{A_2} \geq 2)$ to have strictly more than one photon in mode $A_1$ and $A_2$ respectively. \\

These probabilities can be bounded experimentally in two steps. In the first step, the probabilities $P(\pm1\pm1 |00)$ and $P(\mp 1 \mp 1 |00)$ of having outcomes $\pm 1$ for the measurement of mode $A_1$ and $A_2$ without displacement ($\alpha=\beta=0$) are determined. They provide the following upper bounds 
\begin{eqnarray}
\nonumber
&& p(0,0) \leq P(+1+1|0,0), \quad p(0,1) \leq P(+1-1|0,0),\\
\nonumber
&& p(1,0) \leq P(-1+1|0,0), \quad p(1,1) \leq P(-1-1|0,0). 
\end{eqnarray}
Second, a measurement similar to an autocorrelation measurement using two detectors after a 50/50 beamsplitter is used to measure the probability to get a twofold coincidence $P_{\text c} (A_{1/2})$ after the beamsplitter for both mode $A_1$ and $A_2.$ These coincidence probabilities provide the upper bounds on the missing elements, that is, 
\begin{eqnarray}
\nonumber
&& p(2,1) \leq p(n_{A_1} \geq 2) \leq 2 P_{\text c} (A_{1}) \\
\nonumber
&& p(1,2) \leq p(n_{A_2} \geq 2) \leq 2 P_{\text c} (A_{2}). 
\end{eqnarray}
Once the detection efficiency is included, one gets the following upper bound 
\begin{align}
& S^\star(\alpha, \beta)  \nonumber \\
&= \langle 0 | \sigma_{\sqrt{\eta} \alpha} |0\rangle  \langle 0 | \sigma_{\sqrt{\eta} \beta} |0\rangle  P(+1+1|0,0) \nonumber \\
&+ \langle 0 | \sigma_{\sqrt{\eta} \alpha} |0\rangle  \langle 1 | \sigma_{\sqrt{\eta} \beta} |1\rangle  P(+1-1|0,0)\nonumber \\
&+ \langle 1 | \sigma_{\sqrt{\eta} \alpha} |1\rangle  \langle 0 | \sigma_{\sqrt{\eta} \beta} |0\rangle  P(-1+1|0,0)\nonumber \\
&+ \langle 1 | \sigma_{\sqrt{\eta} \alpha} |1\rangle  \langle 1 | \sigma_{\sqrt{\eta} \beta} |1\rangle  P(-1-1|0,0)\nonumber \\
&+ 2 |\langle 0 | \sigma_{\sqrt{\eta} \alpha} |1\rangle  \langle 1 | \sigma_{\sqrt{\eta} \beta} |0\rangle  | \nonumber \\
&\ \ \ \times \min \{  \sqrt{P(+1+1|0,0) P(-1 -1 |0,0)},  \nonumber \\
& \ \ \ \ \ \ \ \ \ \ \ \ \ \ \ \sqrt{P(+1-1|0,0) P(-1+1|0,0)} \}  \nonumber \\
&+ 2\sqrt{2}|\langle 1 | \sigma_{\sqrt{\eta} \alpha} |0\rangle  \langle 1 | \sigma_{\sqrt{\eta} \beta} |2\rangle  | \nonumber \\
&\ \ \ \times \min\{ \sqrt{ P_c (A_2)P(+1-1|0,0)}, \nonumber \\
& \ \ \ \ \ \ \ \ \ \ \ \ \ \ \ \ \ \ \ \ \sqrt{ P_c (A_2) P(-1-1|0,0)} \}  \nonumber \\
&+ 2\sqrt{2}|\langle 1 | \sigma_{\sqrt{\eta} \alpha} |2\rangle  \langle 1 | \sigma_{\sqrt{\eta} \beta} |0\rangle  | \nonumber \\
&\ \ \ \times \min\{ \sqrt{ P_c (A_1) P(-1+1|0,0)},\nonumber \\
& \ \ \ \ \ \ \ \ \ \ \ \ \ \ \ \ \ \ \ \ \sqrt{ P_c (A_1)P(-1-1|0,0)} \}  \nonumber \\
\label{Sstar}
&+ 2 P_c (A_1) + 2 P_c (A_2),
\end{align}
where
\begin{align}
\nonumber
\langle 0 | \sigma_{\sqrt{\eta} \alpha} |0\rangle &= -1 + 2 e^{-({\sqrt{\eta} \alpha})^2}\\
\nonumber
\langle 1 | \sigma_{\sqrt{\eta} \alpha} |1\rangle &= -1 + 2 ({\sqrt{\eta} \alpha})^2 e^{-({\sqrt{\eta} \alpha})^2}\\
\nonumber
\langle 0 | \sigma_{\sqrt{\eta} \alpha} |1\rangle &= -2 ({\sqrt{\eta} \alpha}) e^{-({\sqrt{\eta} \alpha})^2}\\
\nonumber
\langle 0 | \sigma_{\sqrt{\eta} \alpha} |2\rangle &=  \sqrt{2} ({\sqrt{\eta} \alpha})^2 e^{-({\sqrt{\eta} \alpha})^2}\\
\nonumber
\langle 1 | \sigma_{\sqrt{\eta} \alpha} |2\rangle &= - \sqrt{2}({\sqrt{\eta} \alpha})^3 e^{-({\sqrt{\eta} \alpha})^2}
\end{align}
In practice, the separable bound is obtained by inserting directly the measured probabilities $P(\pm1\pm 1|0,0),$ $P(\mp 1 \pm1|0,0),$ $P_{\text c} (A_{1})$ and $P_{\text c} (A_{2})$ into the previous expression. Note that there is no need to know the amplitude of the displacement and the detector efficiency separately as only the knowledge of the product $\sqrt{\eta} \alpha$ is needed. This is convenient as $\alpha^2 \eta$ can be directly obtained from the click rate on the detector. \\

Finally, the mean value $Q(\alpha,\beta)$ of $M(\alpha, \beta)$ is measured. If there is a value for the couple $\alpha, \beta$ such that $Q(\alpha,\beta) - S^\star(\alpha, \beta) >0,$ we deduce that the assumption on separability does not hold, that is, the photonic modes $A_1$ and $A_2$ are entangled. Since the state describing $A_2$ is obtained from a local operation on the phononic state, $Q(\alpha,\beta) - S^\star(\alpha, \beta) >0$ also certifies entanglement between the photon mode $A_1$ and the phonon mechanical mode.\\ 

\section{Appendix C - Estimation of the experimental value of the witness observable}
We here estimate the mean value of $M(\alpha, \beta)$ that can be obtained in practice, that is, we estimate the value of $Q(\alpha,\beta)$ using a realistic model of the proposed experiment. We consider the case where the mechanical oscillator is not exactly prepared in its ground state at the beginning of the experiment but has a main thermal excitation $n_0.$ The corresponding state can be written as a mixture of coherent states $|\gamma \rangle,$ that is 
\begin{equation}
\rho_b= \frac{1}{\pi n_0}\int d^2\gamma \-\ e^{-|\gamma|^2/n_0} |\gamma \rangle \langle \gamma |.
\end{equation}
We then consider that the blue-detuned excitation is on during a time interval $T_1$ such that the probability that at least one photon-phonon pair is created is given by $p=1 - e^{-2\tilde{g}_+ T_1}.$ The red-detuned excitation is then switched on during a time interval $T_2$ such that the phonon-photon conversion efficiency is given by $T=1-e^{-2\bar{g}_- T_2}.$ 
The detection efficiency is $\eta$ for both mode $A_1$ and $A_2$. Following the procedure presented in Ref. \cite{Caprara16}, we find

\begin{widetext}
\begin{align}
\label{jointprob}
&P(+1+1|\alpha, \beta)  \nonumber \\
=& \frac{1-p}{1+n_0 \eta T - p(-1+\eta  + n_0 \eta)(-1+\eta T) } \times
e^{-\Big[ \frac{
\eta |\alpha|^2 [(1 + p (-1 + \eta T) + n_0 \eta T )]     
+\eta |\beta|^2 [(1 + p (-1 + \eta + n_0 \eta)  )]
+\eta^2(\alpha \beta + \alpha ^* \beta^*) (1 + n_0) \sqrt{pT} 
}
{1+n_0 \eta T - p(-1+\eta  + n_0 \eta)(-1+\eta T)}\Big]}, 
\end{align}

\end{widetext}
\begin{align}
 \label{localAprob}
P(+1|\alpha) = (1-p)\frac{e^{-\frac{\eta |\alpha|^2(1-p)}{p (\eta +\eta  n_0-1)+1}}}{p(\eta + \eta n_0-1)+1},\\
\label{localBprob}
P(+1|\beta) = (1-p)\frac{e^{-\frac{ \eta |\beta|^2 (1-p) }{\eta  T (n_0+p)-p+1}} }{\eta  T (n_0+p)-p+1}.
\end{align} 
These expressions allows us to deduce the expected mean value for the witness observable $Q(\alpha, \beta)$ using 
\begin{align}
\nonumber
Q(\alpha, \beta) = 1 - 2P(+1|\alpha) - 2P(+1|\beta)+ 4 P(+1+1|\alpha, \beta).
\end{align}
This observed value $Q(\alpha, \beta)$ must then be compared to the maximum value $S^\star{(\alpha, \beta)}$ for all separable states to assess the presence of entanglement. To estimate $S^\star(\alpha, \beta)$, we still need to estimate the probabilities of coincidence counts after a 50/50 beamsplitter on each mode. For example, sending mode $A_1$ into a 50-50  beamsplitter yields two output modes $a_1$ and $a'_1$ with photon number probabilities
\begin{align}
&P(n_{a_1}=m,n_{a'_1}=0) \nonumber \\
=& 2^{-m}(1-p) \frac{[p(1 + n_0)\eta]^m}{[1-p(1-\eta-n_0 \eta)]^{m+1}} \nonumber \\
=&P(n_{a_1}=0,n_{a'_1}=m)  \nonumber ,
\end{align}
which then allows us to obtain the probabilities for coincidence counts on mode $A_1$,
\begin{align}
P_c(A_1) &= 1 -P(n_{a_1}=0,n_{a'_1}=0) \nonumber \\
&\quad - \sum_{m=1} ^\infty P(n_{a_1}=m,n_{a'_1}=0)  \nonumber \\
&\quad - \sum_{m=1} ^\infty P(n_{a_1}=0,n_{a'_1}=m)  \nonumber  \\
&=1 - \frac{1-p}{1-p(1-\eta-n_0 \eta)}  \nonumber \\
\nonumber
&- 2\Bigg[\frac{(1+n_0)(1-p)\eta p}{(2-p(2-\eta-\eta n_0))(1+p(-1+\eta+n_0 \eta))}\Bigg].   
\end{align}
Similarly for mode $A_2,$ the probability for coincidence counts is given by
\begin{align}
&P_c(A_2)  \nonumber \\
&= 1 - \frac{1-p}{1-p+(n_0 + p)T \eta }  \nonumber \\
\nonumber
&-2\Bigg[\frac{(1-p)(n_0 + p)T\eta}{(2+n_0 T\eta +p(-2+T\eta)   ) (1+n_0 T\eta +p(-1+T\eta)   )}\Bigg].
\end{align}

\begin{figure}[h]

\includegraphics[width=8.0 cm,trim=4.2cm 3.5cm 4.2cm 3.5cm]{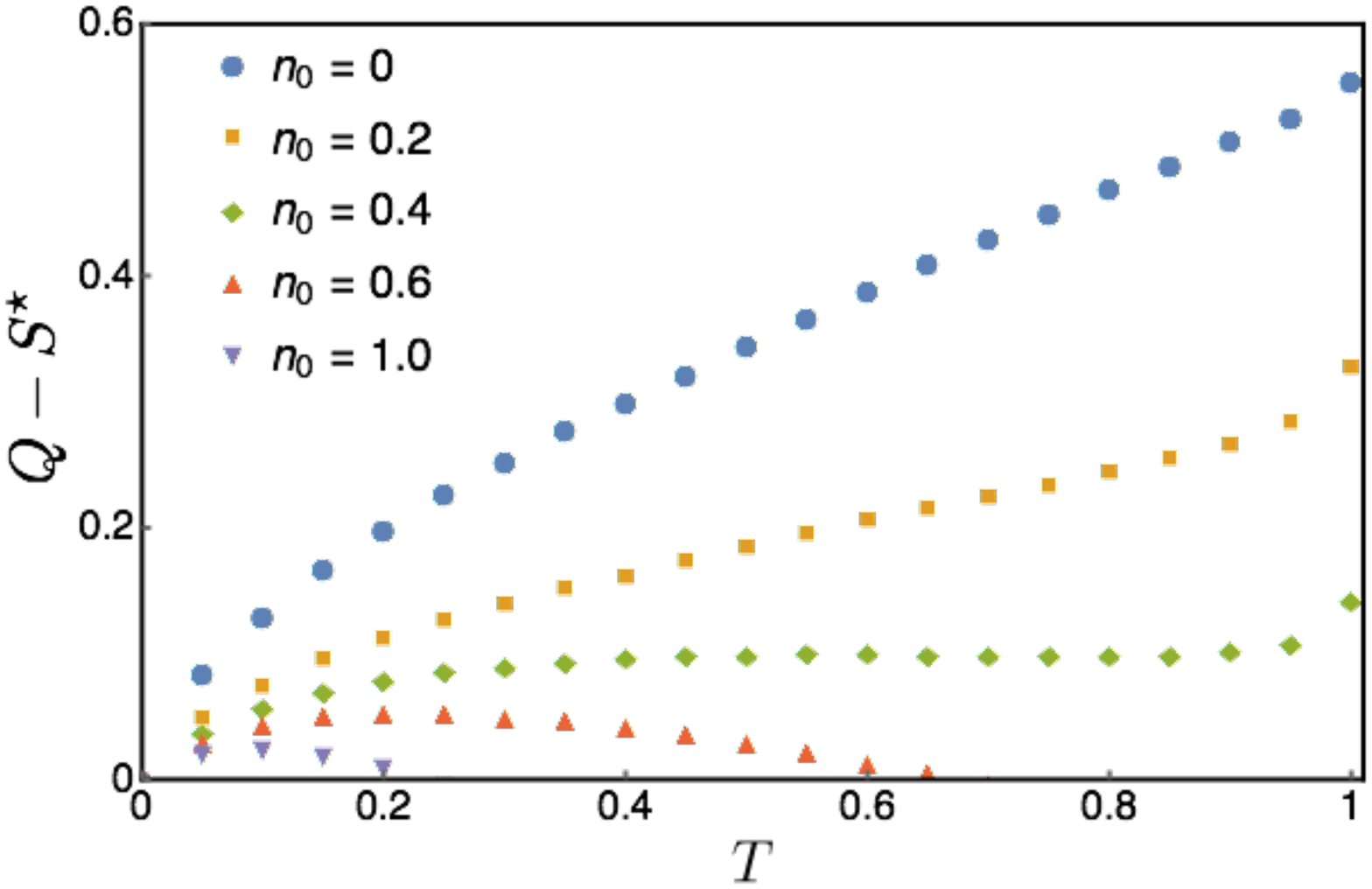}

\caption{Optimal values of the witness observable $Q$ with respect to the separable bound $S^\star$ for unit detection efficiency $(\eta=1)$ as a function of the phonon-photon conversion efficiency $T = 1 - e^{-2\bar{g}_- T_2}$ for various values of the mechanical thermal noise $n_0.$ The difference $Q-S^\star$ is optimised over the displacement amplitudes $\alpha, \beta$ and the amount of initial squeezing $\bar{g}_+ T_1$.}
\label{resistancetothermal}
\end{figure}

These probabilities allow us to estimate the value of $S^\star(\alpha, \beta)$ from Eq. \eqref{Sstar} and thus to deduce $Q(\alpha, \beta) - S^\star(\alpha, \beta).$ The result is shown in Fig. 2 of the Main Text as a function of the conversion efficiency $T$ for various detection efficiency in the case where the mechanical system is initially in its ground state $n_0=0.$ Fig. \ref{resistancetothermal} shows $Q(\alpha, \beta) - S^\star(\alpha, \beta)$ for various values of initial thermal excitations in the resonator $n_0.$

It is important to mention that the same values for $Q-S^\star$ are obtained in the case where $\eta$ corresponds to the efficiency with which the photons are generated and transmitted until the displacement operation and are subsequently detected with unit efficiency detectors, although with different displacement amplitudes. This is clear mathematically since the same statistics are obtained with loss operating before or after the displacement operations provided that the displacement amplitude is changed accordingly, c.f. section A. The results presented in Fig. 2 of the main text can thus be seen as the expected value of the observable witness with respect to the separable bound for various overall detection efficiencies, including all the loss from the generation to the detection of photons. 

\section{Appendix D - Statistical Analysis}

What we have calculated so far for expected values of $Q-S^\star$  are asymptotic values. These values are derived from probabilities computed using the Born rule, and are guaranteed to be the observed quantities only in the situation where the number of experimental runs tends to infinity.

The number of runs available, however, are limited in practice. This would lead to a scenario where, due to statistical fluctuations, an experiment that reveals $Q-S^\star>0$  in the asymptotic case might not reveal this entanglement with limited runs.
To overcome this, we require sufficient runs for an estimator $\overline{Q(\alpha, \beta)-S^\star(\alpha, \beta)}$, so that its variance will be low as compared to its asymptotic value $Q(\alpha, \beta)-S^\star(\alpha, \beta)$.
This guarantees that a significant violation is likely to be experimentally observed. Once such a violation is observed, a similar calculation could be made to guarantee that the observed statistics are not compatible with an entangled state, i.e. bounding the possible p-value.

We now present how we form an appropriate estimator for our witness. As an initial example, an event with probability $P(m)$ can be estimated with $N$ runs, using a sample estimator $\overline{P(m)}$
\begin{align}
\overline{P(m)} = \frac{1}{N}\sum_i^N x_i \ \  , \  x_i \left\{ \begin{array}{ll}
            +1 &\text{if the $i$-th run shows `m'} \\
            0  &\text{otherwise}
        \end{array} \right.,   \nonumber 
\end{align}

This is a consistent estimator, as the expectation value of the sample gives the quantity to be estimated
\begin{align}
\mathbb{E}\Big ( \overline{P(m)} \Big) &= \mathbb{E} \Big( \frac{1}{N}\sum_i^N x_i \Big) \nonumber \\
&= \frac{1}{N}\sum_i^N   \mathbb{E}(x_i)  \nonumber \\
&= P(m).
\end{align}

The variance of the sample estimator with $N$ runs can be found to scale with $\frac{1}{N}$, since
noting that $x_i^2 = x_i$, we have 
\begin{align}
\mathbb{E} \Big( [\overline{P(m)} ]^2 \Big) &= \mathbb{E} \Big( \frac{1}{N^2} \sum_{i,j}^N x_i x_j \Big)  \nonumber \\
&= \frac{1}{N^2} \mathbb{E} \Big ( \sum_{i=j} x_i x_j + \sum_{i \neq j} x_i x_j  \Big )  \nonumber \\
&=  \frac{1}{N^2} \Big [ \sum_{i=j} \mathbb{E}( x_i x_j) + \sum_{i \neq j} \mathbb{E}( x_i x_j)  \Big]  \nonumber \\
&= \frac{1}{N^2} \Big [ \sum_{i}^N \mathbb{E}( x_i ^2) + \sum_{i }^N\mathbb{E}( x_i)  \sum_{j\neq i }\mathbb{E}( x_j)  \Big]  \nonumber  \\
&= \frac{1}{N^2} \Big[ N P(m) + N(N-1) P(m)^2 \Big] ,
 \end{align}
and we compute
\begin{align}
\text{Var}\Big( \overline{P(m)} \Big) &= \mathbb{E}\Big( \Big[\overline{P(m)}\Big]^2 \Big) -\mathbb{E}\Big( \overline{P(m)} \Big) ^2 \nonumber \\
&= \frac{1}{N}P(m)\big[ 1-P(m) \big],
\end{align}
 allowing us to assess the variance of $P(m)$ as a function of the probability $P(m)$ and the number of runs. 
If the variance of the sample estimator is small using $N$ runs, one can be confident that in those $N$ runs the sample estimator gives a value close to its expectation value. 

Let us now move on to more complicated combinations of estimators, starting with linear combinations. The variance of linear combinations of probabilities of different events can be easily computed if the runs are uncorrelated. For example, if we estimate $P(m)$ from a sample of $N_1$ runs, and $ P(n)$ from a sample of separate $N_2$ runs, then 
\begin{align}
\text{Var} \Big( \alpha \overline{ P(m)} +  \beta \overline{ P(n)}  \Big)  = \alpha^2 \text{Var}\Big( \overline{ P(m)} \Big) + \beta^2 \text{Var} \Big( \overline{ P(n)}  \Big)
\end{align}
by the well known variance addition formula for uncorrelated data. 

Other examples include the product of two probabilities, such as  $P(m)P(n)$. One could use a natural choice for the estimator, where probabilities $P(m) (P(n))$ are assessed with $N_A (N_B)$ runs over separate data, such that
\begin{align}
\overline{P(m)  P(n)} = \frac{1}{N_A}\sum_i^{N_A} x_i \frac{1}{N_B}\sum_j^{N_B} y_j,
\end{align}
where $y_i$ is assigned to indicate runs showing `n', similar to the case for $P(m)$. One can easily see that our estimator is consistent, giving $\mathbb{E}\Big( \overline{P(m)  P(n)} \Big) = P(m) P(n)$. One can also get the variance of this chosen estimator as an explicit function of $P(m),P(n), N_A$ and $N_B$.

Considering now the quantites in $S^\star$, however, we have terms involving the square root of products of two probabilities. 
Choosing a natural choice for the estimator as we did before does not give us the analytical functions for the variance we would like, since if for $\sqrt{P(m) P(n)}$ we use the estimator
\begin{align}
\overline{\sqrt{P(m) P(n)}} = \sqrt{\frac{1}{N_A}\sum_i^{N_A} x_i \frac{1}{N_B}\sum_j^{N_B} y_j
},
\end{align}
then we obtain 

\begin{align}
\mathbb{E}\Big(\Big[ \overline{\sqrt{P(m) P(n)}} \Big] ^2\Big) = P(m)P(n) 
\end{align} but  

\begin{align}
\mathbb{E}\Big( \overline{\sqrt{P(m) P(n)}} \Big) =& \mathbb{E}  \Bigg( \sqrt{   \frac{1}{N_A}\sum_i^{N_A} x_i \frac{1}{N_B}\sum_j^{N_B} y_j}   \Bigg) \nonumber  \\
\leq&   \sqrt{\mathbb{E}\Big( \frac{1}{N_A}\sum_i^{N_A} x_i \frac{1}{N_B}\sum_j^{N_B} y_j} \Big)  \nonumber  \\
=& \sqrt{P(m) P(n)},
\end{align}
leading only to the trivial bound $\text{Var}\Big(\overline{\sqrt{P(m) P(n)}} \Big)\geq0$. We thus do not consider such estimators directly in our variance assessment for nonlinear terms.

We instead consider a linearisation on the nonlinear quantities in $S^\star(\alpha, \beta)$ by finding the tangent surface at a point. 
We recall that for a 2-dimensional function $f(x,y)$, the tangent surface at the point $(x_0, y_0)$ is given by 
\begin{align}
&f_{linear}  \nonumber \\
=& f(x_0, y_0) + f_x (x_0, y_0) (x-x_0)  + f_y (x_0, y_0) (y-y_0) , \nonumber  
\end{align}
where the partial derivatives of $f(x,y)$ to $x(y)$ are $f_{x(y)} (x,y) $ respectively.

With this, we can find the tangent surface to each of the square root terms  and obtain linear combinations of probabilities. For example, for $z = \sqrt{P(+1+1|0,0)\ P(-1-1|0,0)}$ at the point $(A,B)$, we have 
\begin{align}
z &= \sqrt{P(+1+1|0,0)\ P(-1-1|0,0)} \nonumber \\
&\leq \frac{1}{2} \Bigg[ \sqrt{\frac{B}{A}} P(+1+1|0,0) + \sqrt{\frac{A}{B}} P(-1-1|0,0) \Bigg] \nonumber \\
&=z_{\text{linear}}.  \nonumber 
\end{align}
In this case it is important to note that the resulting linear combination of probabilities forms an upper bound due to the concavity of the square root function, overestimating $z$ in a conservative manner. We stress that any surface $z_{\text{linear}}$ with nonzero $A$ and $B$ is a valid upper bound on $z$. Since we overestimate quantities that are in $S^\star(\alpha, \beta)$, this does not lead to a false conclusion of entanglement. To assess the value of $z_{\text{linear}}$, we can now use individual estimators $\overline{P{(+1+1|0,0)}}$  and $\overline{P{(-1-1|0,0)}}$ that converge to $P(+1+1|0,0)$ and $P(-1-1|0,0)$ respectively, so that
\begin{align}
\overline{z_{linear}  }= \frac{1}{2} \Bigg[ \sqrt{\frac{B}{A}} \overline{P(+1+1|0,0)} + \sqrt{\frac{A}{B}}\overline{ P(-1-1|0,0) }\Bigg], \nonumber
\end{align}
and whose variance we can easily compute assuming separate runs in the estimation for each term.

At this point, $A$ and $B$ can independently take any nonzero value from 0 to 1, and still give valid linearised upper bounds. To select more optimal values for $A$ and $B$ in such a linearised estimator, one can perform an initial calibration experiment. Given these values, one can then form a valid, but close to optimal $z_{\text{linear}}$. We first point out with an infinite number of runs for the calibration, an accurate calibration is possible, yielding $A = P(+1+1|0,0)_{\text{cal}} \rightarrow P(+1+1|0,0)$ and $B = P(-1-1|0,0)_{\text{cal}} \rightarrow P(-1-1|0,0)$. Furthermore, consistent estimators asymptotically converge to the quantum values, giving $\overline{P(+1+1|0,0)} \rightarrow P(+1+1|0,0)$ and $\overline{P(-1-1|0,0)} \rightarrow P(-1-1|0,0)$. Therefore in the case where $P(+1+1|0,0)$ and $P(-1-1|0,0)$ are nonzero, the asymptotic $z_{\text{linear}} = z$.

We thus create an estimator   $\overline{{Q(\alpha, \beta)-S^\star_{\text{linear}}(\alpha, \beta)} }$ in this best case scenario, made up of a linear combination of individual estimators so that we can easily assess its variance
\begin{align}
 &\overline{Q(\alpha, \beta) - S^\star(\alpha, \beta)}  \leq \overline{Q(\alpha, \beta) - S^\star_{\text{linear}}(\alpha, \beta) }\nonumber \\
=& \overline{P(+1+1| \alpha, \beta)} + \overline{P(+1-1| \alpha, \beta)}  \nonumber  \\
+& \overline{ P(-1+1| \alpha, \beta)} + \overline{P(-1-1| \alpha, \beta)}  \nonumber \\
-&\langle 0 | \sigma_{\sqrt{\eta} \alpha} |0\rangle  \langle 0 | \sigma_{\sqrt{\eta} \beta} |0\rangle \overline{P(+1+1|0,0)} \nonumber \\
-&\langle 0 | \sigma_{\sqrt{\eta} \alpha} |0\rangle  \langle 1 | \sigma_{\sqrt{\eta} \beta} |1\rangle \overline{P(+1-1|0,0)}  \nonumber  \\
-&\langle 1 | \sigma_{\sqrt{\eta} \alpha} |1\rangle  \langle 0 | \sigma_{\sqrt{\eta} \beta} |0\rangle \overline{P(-1+1|0,0)}   \nonumber  \\
-&\langle 1 | \sigma_{\sqrt{\eta} \alpha} |1\rangle  \langle 1 | \sigma_{\sqrt{\eta} \beta} |1\rangle \overline{P(-1-1|0,0)}   \nonumber \\
-& |\langle 0 | \sigma_{\sqrt{\eta} \alpha} |1\rangle  \langle 1 | \sigma_{\sqrt{\eta} \beta} |0\rangle  |\nonumber  \\
 \times& \text{min} 
 \Big [  k_1 \  \overline{P(+1+1|0,0)} + k_1^{-1}\overline{P(-1-1|0,0)} , \nonumber \\
 &\ \ \ \ \ \   k_2 \  \overline{P(+1-1|0,0)} + k_2^{-1}\overline{P(-1+1|0,0)} \Big]  \nonumber  \\
-& \sqrt{2}|\langle 1 | \sigma_{\sqrt{\eta} \alpha} |0\rangle  \langle 1 | \sigma_{\sqrt{\eta} \beta} |2\rangle  |  \nonumber  \\
 \times& \text{min} 
 \Big [  k_3 \  \overline{P_c(A_2)} + k_3^{-1}\overline{P(+1-1|0,0)} , \nonumber \\
 &\ \ \ \ \ \   k_4 \   \overline{P_c(A_2)} + k_4^{-1}\overline{P(-1-1|0,0)} \Big]  \nonumber  \\
-& \sqrt{2}|\langle 1 | \sigma_{\sqrt{\eta} \alpha} |2\rangle  \langle 1 | \sigma_{\sqrt{\eta} \beta} |0\rangle  |  \nonumber  \\
 \times& \text{min} 
 \Big [  k_5 \  \overline{P_c(A_1)} + k_5^{-1}\overline{P(-1+1|0,0)} , \nonumber \\
 &\ \ \ \ \ \   k_6 \   \overline{P_c(A_1)} + k_6^{-1}\overline{P(-1-1|0,0)} \Big]  \nonumber  \\
-& 2\Big [\overline{ P_c(A_1) }+ \overline{P_c(A_2) }\Big] ,
\end{align}
where\footnote{If one has some knowledge of the state parameters, one might even compute the expected asymptotic values to use as the calibration parameters. The use of this knowledge does not affect the validity of the entanglement conclusion from the actual experiment, as it only varies the overestimation of each term of the witness.}
 $k_1 =\sqrt{\frac{P(-1-1|0,0)_{\text{cal}}}{P(+1+1|0,0)_{\text{cal}}}} $, $k_2 =\sqrt{\frac{P(-1+1|0,0)_{\text{\text{cal}}}}{P(+1-1|0,0)_{\text{\text{cal}}}}} $, $k_3 =\sqrt{\frac{P(+1-1|0,0)_{\text{cal}}}{P_c(A_2)_{\text{cal}}}} $, $k_4 =\sqrt{\frac{P(-1-1|0,0)_{\text{cal}}}{P_c(A_2)_{\text{cal}}}} $, $k_5 =\sqrt{\frac{P(-1+1|0,0)_{\text{cal}}}{P_c(A_1)_{\text{cal}}}} $ and $k_6 =\sqrt{\frac{P(-1-1|0,0)_{\text{cal}}}{P_c(A_1)_{\text{cal}}}} $.

With a budget of $N_\text{total}$ runs, we can now minimise the variance of $\overline{Q(\alpha, \beta) - S^\star_{\text{linear}}(\alpha, \beta) } $ over possible distributions of $N_\text{total}$ runs across each estimator term within.
For our purposes, we will consider a number of runs $N_{\text{total}}$ sufficient for revealing entanglement if the variance for an accurately calibrated $\overline{Q(\alpha, \beta) - S^\star_{\text{linear}} (\alpha, \beta)} $ is such that
\begin{align}
&\sqrt{\text{Var}\Big( \overline{Q(\alpha, \beta) - S^\star_{\text{linear}}(\alpha, \beta) }  \Big) }  \nonumber \\
&\leq \frac{1}{3} \big [ Q(\alpha, \beta)-S^\star(\alpha, \beta) \big].
\end{align}
 In the case discussed in the Main Text, where one has $\eta = 10\%$, $n_0 = 0.2$ and a state-swap efficiency of $T=30\%$ we find that $N_{\text{total}}=750000$ runs are sufficient.
 
\section{Appendix E - Implementation}

\begin{figure}[h]
\begin{center}

\includegraphics[width=8.5 cm,trim=3.5cm 3cm 3.5cm 3cm,clip ]{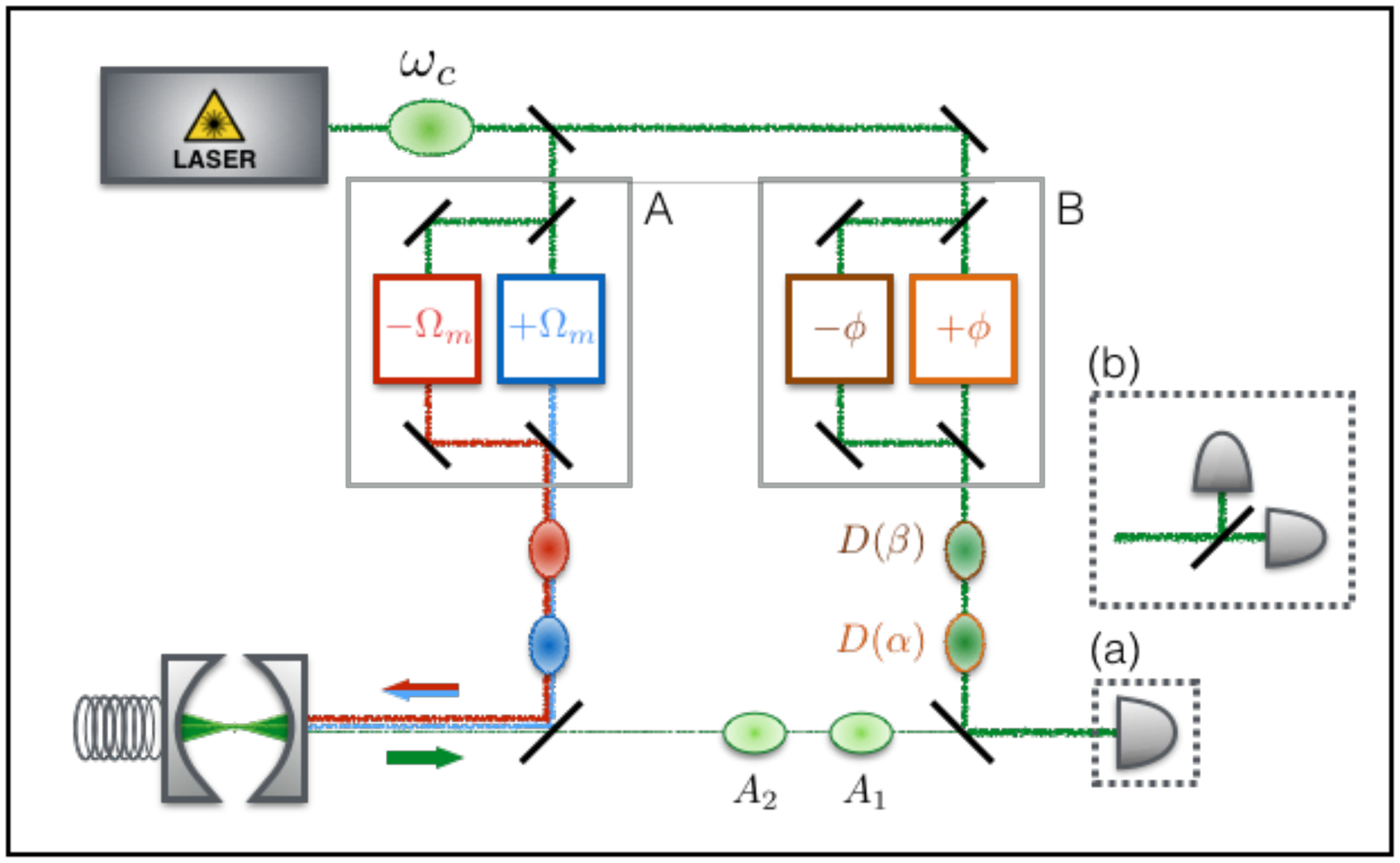}

\end{center}
\caption{Schematic of a possible setup. A laser set to the resonant frequency of the cavity is sent onto a beamsplitter where the beam is diverted towards interferometers A and B. Interferometer A is set up such that the pulse on the shorter/longer path is respectively detuned by plus/minus the mechanical frequency. The blue and red detuned pulses then enter the cavity and result in optical modes $A_1$ and $A_2$ which we detect for the witness.  Interferometer B separates the incoming light into two paths, each modified so as to obtain the amplitude and phase of the respective displacement operation. Light leaving interferometer B then combines with the light from the cavity in order to realise the displacement operations before the photon detection. The two methods of detection required are shown in (a) using a single detector capable of measuring both $A_1$ and $A_2$, and in (b) with two detectors after a beamsplitter to measure coincidences. }
\label{schematic}
\end{figure}

 We present in Fig. \ref{schematic} a possible way to implement our proposal. Pulses which are created at the cavity frequency $\omega_c,$ are split before being sent into Mach-Zehnder interferometers. The pulses in the first interferometer are used to  drive the opto-mechanical system, that is, the frequency in the short and long arm is shifted so as to be resonant with the relevant optomechanical sidebands implementing the two-mode squeezing and phonon-photon state transfer operations respectively. Each arm of the second interferometer is equipped with phase and amplitude modulators to set the the phase and amplitude of displacement operations. The latter is indeed implemented by combining the state to be displaced and a coherent state into a partially reflecting beamsplitter. The modulators are here to guarantee that both states are indistinguishable in all degrees of freedom, differing only with regards to their photon number distribution.

We point out that this scheme imposes minimal requirements on the phase stability. As can be seen in Fig. \ref{schematic}, the only phase stability requirement is that the relative path length fluctuations in each of the interferometers A and B are small with respect to the wavelength. Phase stability in the larger interferometer that separates the initial laser pulses towards interferometers A and B is not required. Furthermore, for the detection of outcomes, at most twofold coincidences are required. 

Finally, let us list the steps one should take to certify optomechanical entanglement. 
\begin{enumerate}
\item Evaluate  $S^\star(\alpha, \beta)$ by first recording the events `click' and `no-click' without displacement (measurement scheme (a)) to obtain the probabilities $ P(\pm1,\pm1|0,0)$ and $P(\pm1,\mp1| 0,0)$; and then introducing a beamsplitter and recording the coincidences (measurement scheme (b)) to obtain  $P_c(A_1)$ and $P_c(A_2)$.
\item Evaluate $Q(\alpha, \beta)$ by recording the events `click' and `no-click' with settings $\alpha$ and $\beta$, to obtain $P(\pm1,\pm1|\alpha,\beta)$ and $P(\pm1,\mp1|\alpha,\beta)$.
\item Conclude entanglement  if there is a couple $(\alpha, \beta)$ such that $Q(\alpha, \beta)-S^\star(\alpha, \beta)>0$.
\end{enumerate}

\end{appendix}


\begin{thebibliography}{3}

\bibitem{Bell64} J.S. Bell, Physics {\bf 1}, 195 (1964)

\bibitem{Hensen15} B. Hensen \textit{et al.} Nature {\bf 526}, 682 (2015)

\bibitem{Rosenfeld17} W. Rosenfeld, \textit{et al.} Phys. Rev. Lett. {\bf 119}, 010402 (2017)

\bibitem{Giustina15} M. Giustina \textit{et al.} Phys. Rev. Lett. {\bf 115}, 250401 (2015)

\bibitem{Shalm15} L.K. Shalm \textit{et al.} Phys. Rev. Lett. {\bf 115}, 250402 (2015)

\bibitem{Caprara16} V. Caprara Vivoli, T. Barnea, C. Galland, and N. Sangouard, Phys. Rev. Lett., {\bf 116}, 070405 (2016)

\bibitem{Hofer16} S.G. Hofer, K.W. Lehnert, and K. Hammerer, Phys. Rev. Lett., {\bf 116}, 070406 (2016)

\bibitem{Asjad18} M. Asjad, J. Manninen, E. Selenius, R. Ojajarvi, P. Kuusela, and F. Massel, arXiv:1803.00331

\bibitem{O'Connell10} A.D. O'Connell, M. Hofheinz, M. Ansmann, R.C. Bialczak, M. Lenander, E. Lucero, M. Neeley, D. Sank, H. Wang, M. Weides, J. Wenner, J.M. Martinis, and A.N. Cleland, Nature {\bf 464}, 697 (2010)

\bibitem{Teufel11} J.D. Teufel, T. Donner, D. Li, J.W. Harlow, M.S. Allman, K. Cicak, A.J. Sirois, J.D. Whittaker, K.W. Lehnert, and R.W. Simmonds, Nature {\bf 475}, 359 (2011)

\bibitem{Chan11} J. Chan, T.P.M. Alegre, A.H. Safavi-Naeini, J.T. Hill, A. Krause, S. Groblacher, M. Aspelmeyer, and O. Painter, Nature {\bf 478}, 89 (2011)

\bibitem{Palomaki13} T.A. Palomaki, J.D. Teufel, R.W. Simmonds, and K.W. Lehnert, Science {\bf 342}, 710 (2013)

\bibitem{Riedinger16} R. Riedinger, S. Hong, R.A. Norte, J.A. Slater, J. Shang, A.G. Krause, V. Anant, M. Aspelmeyer, and S. Groblacher, Nature {\bf 530}, 313 (2016)

\bibitem{Hong17} S. Hong, R. Riedinger, I. Marinkovic, A. Wallucks, S. G. Hofer, R. A. Norte, M. Aspelmeyer, and S. Groblacher, Science {\bf 358}, 203 (2017)

\bibitem{Riedinger17} R. Riedinger, A. Wallucks, I. Marinkovic, C. Loschnauer, M. Aspelmeyer, S. Hong, and S. Groblacher, arXiv:1710.11147

\bibitem{Kimble08} H.J. Kimble, Nature {\bf 453}, 1023 (2008)

\bibitem{Borkje11} K. Borkje, A. Nunnenkamp, and S.M. Girvin, Phys. Rev. Lett. {\bf 107}, 123601 (2011)

\bibitem{Galland14} C. Galland,  N. Sangouard, N. Piro, N. Gisin, and T. J. Kippenberg, Phys. Rev. Lett. {\bf 112}, 143602 (2014)

\bibitem{Scarani12} V. Scarani, Acta Physica Slovaca {\bf 62}, 347 (2012)

\bibitem{Acin06} A. Acin, N. Gisin, and L. Masanes, Phys. Rev. Lett. {\bf 97}, 120405 (2006)

\bibitem{Lydersen10} L. Lydersen, C. Wiechers, C. Wittmann, D. Elser, J. Skaar, and V. Makarov, Nature Photonics {\bf 4}, 686 (2010)

\bibitem{Sekatski18} P. Sekatski, J.-D. Bancal, S. Wagner, and N. Sangouard, arXiv:1802:04163

\bibitem{CHSH69} J.F. Clauser, M.A. Horne, A. Shimony, and R.A. Holt, Phys. Rev. Lett. {\bf 23}, 880 (1969)

\bibitem{Cohen15} J.D. Cohen, S.M. Meenehan, G.S. MacCabe, S. Groblacher, A.H. Safavi-Naeini, F. Marsili, M.D. Shaw, and O. Painter, Nature {\bf 520}, 522 (2015)

\bibitem{Hofer11} S. G. Hofer, W. Wieczorek, M. Aspelmeyer, and K. Hammerer, Phys. Rev. A., {\bf 84}, 052327 (2011)

\bibitem{Vanner13} M.R. Vanner, M. Aspelmeyer, and M.S. Kim, Phys. Rev. Lett. {\bf 110}, 010504 (2013)

\bibitem{Aspelmeyer10} M. Aspelmeyer, S. Groeblacher, K. Hammerer, and N. Kiesel, J. Opt. Soc. Am. B {\bf 27}, A189-A197 (2010)

\bibitem{Kuzmich00} A. Kuzmich I. A. Walmsley, and L. Mandel, Phys. Rev. Lett. {\bf 85}, 1349 (2000)

\bibitem{Lee09} S.-W. Lee, H. Jeong, and D. Jaksch, Phys. Rev. A {\bf 80}, 022104 (2009)

\bibitem{Brask12} J. Bohr Brask and R. Chaves, Phys. Rev. A {\bf 86}, 010103(R) (2012)

\bibitem{Caprara15} V. Caprara Vivoli, P. Sekatski, J.-D. Bancal, C.C.W. Lim, A. Martin, R.T. Thew, H. Zbinden, N. Gisin, and N. Sangouard, New J. Phys. {\bf 17}, 023023 (2015)

\bibitem{Peres96} A. Peres, Phys. Rev. Lett., {\bf 77}, 1413 (1996)

\bibitem{Horodecki96} M. Horodecki, P. Horodecki, and R. Horodecki, Phys. Lett. A , {\bf 23}, 1 (1996)

\bibitem{Chan12} J. Chan, A.H. Safavi-Naeini, J.T. Hill, S. Meenehan, and O. Painter, Applied Physics Letters {\bf 101}, 081115 (2012)

\bibitem{Kuramochi10} E. Kuramochi, H. Taniyama, T. Tanabe, K. Kawasaki, Y.-G. Roh, and M. Notomi, Opt. Express {\bf 18}, 15859 (2010)

\bibitem{Sun13} X. Sun, X. Zhang, C. Schuck, and H. X. Tang, Sci. Rep. {\bf 3}, 1436 (2013)

\end{thebibliography}
\end{document}